\documentclass[prb,twocolumn,superscriptaddress,preprintnumbers,a4,showpacs,floatfix]{revtex4}
\usepackage{amsmath, amssymb, psfrag, tabls, dcolumn, bm, color}
\usepackage[sort&compress]{natbib}
\usepackage{graphicx}
\usepackage{textcomp}

\newcommand\smallurl[1]{{\tiny \url{#1}}}

\newcommand{\be}{\begin{equation}}
\newcommand{\ee}{\end{equation}}
\newcommand{\bea}{\begin{equation*}}
\newcommand{\eea}{\end{equation*}}
\newcommand{\ba}{\begin{array}}
\newcommand{\ea}{\end{array}}
\newcommand{\beqa}{\begin{eqnarray}}
\newcommand{\eeqa}{\end{eqnarray}}
\newcommand{\beqaa}{\begin{eqnarray*}}
\newcommand{\eeqaa}{\end{eqnarray*}}

\newcommand{\matr}{\left( \begin{array}}
\newcommand{\ematr}{\end{array} \right)}

\newcommand{\lsim}{{\;\raise0.3ex\hbox{$<$\kern-0.75em\raise-1.1ex\hbox{$\sim$}}
\;}}
\newcommand{\gsim}{{\;\raise0.3ex\hbox{$>$\kern-0.75em\raise-1.1ex\hbox{$\sim$}}
\;}}

\usepackage{mdwlist}

\def\bcols{\begin{columns}}
\def\ecols{\end{columns}}
\def\bcol{\begin{column}}
\def\ecol{\end{column}}
\def\bit{\begin{itemize}}
\def\eit{\end{itemize}}
\def\ben#1{\begin{enumerate}[#1]}
\def\een{\end{enumerate}}
\def\colb#1{\begin{columns}\begin{column}{#1}}

\def\cole{\end{column}\end{columns}}

\usepackage{color}
\usepackage{framed}
\usepackage{comment}
\definecolor{shadecolor}{gray}{0.875}
\specialcomment{answer}{\begin{shaded}}{\end{shaded}}
\specialcomment{vastaus}{}{}
\specialcomment{pikavastaus}{}{}

\RequirePackage{ifthen}
\newboolean{isvastaus}
\setboolean{isvastaus}{false}

\ifthenelse{\boolean{isvastaus}}
{}
{\excludecomment{vastaus}
}

\begin{document}

\title{Electromechanics of Twisted Graphene Nanoribbons}

\author{Pekka Koskinen}
\address{Department of Physics, NanoScience Center, University of Jyv\"askyl\"a, 40014 Jyv\"askyl\"a, Finland}
\email{pekka.koskinen@iki.fi}



\date{\today}

\begin{abstract}
Graphene nanoribbons are the flimsiest material systems in the world, and they get readily distorted. Distortion by twisting, for one, is important because it couples to ribbon's electronic properties. In this Letter, using simulations with density-functional tight-binding and revised periodic boundary conditions, I show that twisting appears almost equivalent to stretching; electronic structures in a given nanoribbon either upon twisting or upon certain stretching are quantitatively similar. This simple equivalence will provide a valuable guideline for interpreting and designing experiments with these flimsy ribbons.
\end{abstract}
\maketitle

The flimsyness of graphene nanoribbons makes perfect experiments and theory comparisons difficult. Both free-standing and supported ribbons are prone to distortions,\cite{kosynkin_nature_09,li_science_08} that affect also electronic structures; uniaxial stretching, for example, induces systematic changes in the fundamental energy gaps.\cite{hod_NL_09,sun_JCP_08}. Supported ribbons are easier to hold even,\cite{lui_nature_09} but at the same time---because of support interactions---can have even more abrupt distortions such as loops and folds.\cite{li_science_08}. Ribbons can be affected by the support interactions also directly.\cite{varchon_PRB_08}

Graphene nanoribbons have high claims for usage as sensors, spintronic devices, and ballistic transistors, but these claims originate often from simulations without defects, disorder or distortions.\cite{abergel_AP_10} Sure enough, distortions themselves, however complex, can be investigated by classical potentials and long, finite ribbons.\cite{bets_NR_09,li_JPD_10} But for electronic structure one needs a quantum approach,\cite{barone_NL_06,topsakal_PRB_10} which for finite ribbons would soon become expensive. Efficient twisting simulations hence require periodic boundary conditions---but beyond the standard translational symmetry.

%
%

In this Letter I show simulations of twisted zigzag (ZGNR) and armchair (AGNR) graphene nanoribbons, using density-functional tight-binding and revised periodic boundary conditions. This computationally efficient approach enabled simulations of infinitely long ribbons with continuously tunable twist. It turned out that, in terms of gap, density of states and optical spectrum, a twisted ribbon with a fixed length can be imitated by a flat ribbon with a certain stretch. If the ribbon length is allowed to relax, however, the electronic structure remains almost unaffected by twisting; both of these phenomena can be explained in terms of average strain.

\begin{figure}[b!]
\includegraphics[width=7cm]{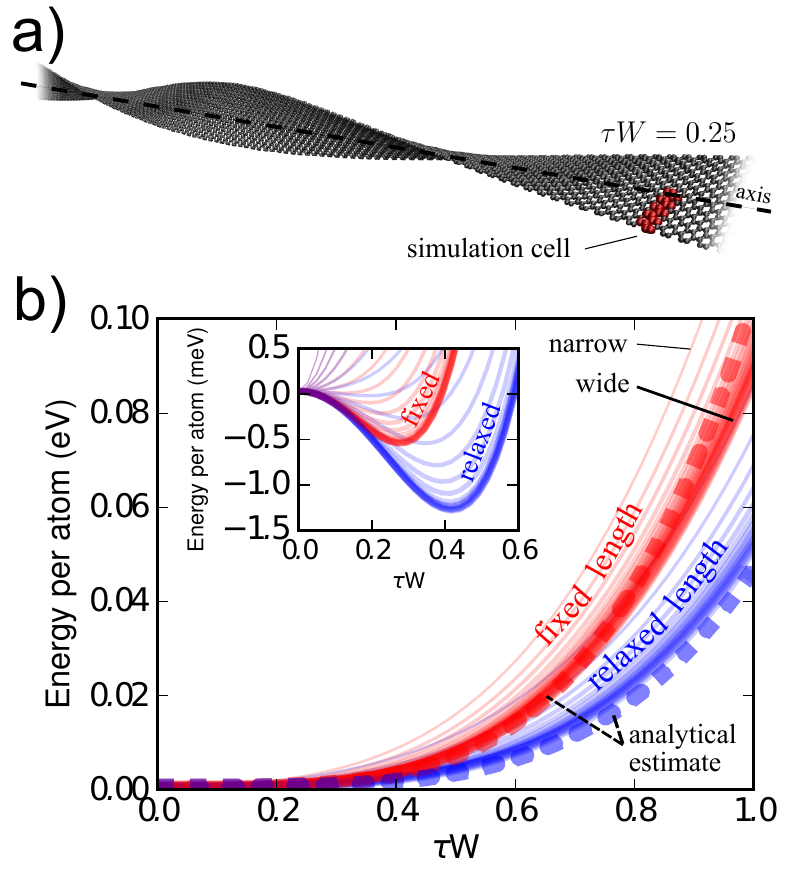}
\caption{(Color online) a) Twisted 20-AGNR with $\tau W=0.25$, together with an illustration of the simulation cell. b) Total energy per atom for twisted, hydrogen-passivated $N$-ZGNRs, either with fixed or relaxed unit cell length; only the narrowest ribbons deviate from the analytical estimates. Inset: energy can decrease upon twisting due to edge stress in unpassivated $N$-AGNRs, and induce spontaneous twisting.}
\label{fig:energy}
\end{figure}

The electronic structure modeling method was density-functional tight-binding, which is computationally efficient and still captures the essentials of graphene's elastic and electronic properties.\cite{koskinen_CMS_09,hotbit_wiki} Spin was not included, leaving magnetic properties outside this work.\cite{kunstmann_PRB_11} I adapted revised periodic boundary conditions to chiral symmetry, in the spirit of Ref.~\onlinecite{mintmire_PRL_92}, and used simulation cells as shown in Fig.\ref{fig:energy}a; this approach is described in Ref.~\onlinecite{koskinen_PRL_10}.

I simulated both bare-edge and hydrogen-passivated $N$-AGNRs and $N$-ZGNRs with $N=5 - 23$, corresponding to widths $W\approx 5 - 45$~\AA. I characterize the twist by the dimensionless parameter $\tau W$, where $\tau$ is the twist angle per unit length; ribbons with the same $\tau W$ have the same strain at the edge and have the visual appearance of being equally twisted. Ribbons were twisted up to $\tau W \sim 1$, which corresponds to ribbons that are twisted full turns when ribbon lengths are $2\pi W$. Atom positions were optimized down to maximum force components of $10^{-4}$~eV/\AA, and the simulation cell lengths were either fixed or fully relaxed.\cite{bitzek_PRL_06,tau_note}

Before turning attention to the electronic structure, it's illustrating to begin by inspecting geometry and energy. The strain in the ribbon at a distance $r$ from the axis is
\begin{equation}
\varepsilon(r) = \varepsilon_z + \frac{(\tau r)^2}{2},
\label{eq:strain}
\end{equation}
where $\varepsilon_z$ is the axial strain (the unit cell strain). For $\varepsilon_z=0$ the edge strain is $(\tau W)^2/8$, or $12.5$~\%\ for $\tau W=1$, and justifies the upper twist limit as a no-tearing condition. For a fixed length ($\varepsilon_z=0$) the elastic energy density $\frac{1}{2}k \varepsilon^2$ yields the total energy per atom the simple expression $E=A_c k/640(\tau W)^4$, where $k=25.5$~eV/\AA$^2$ is the in-plane modulus and $A_c=2.62$~\AA$^2$ is the area per carbon atom; for a relaxed length the corresponding expression is $E=A_c k/1440 (\tau W)^4$. Figure \ref{fig:energy}b, showing the energy for hydrogen-passivated ZGNRs, reveals that the strain really is the main component in energy, rendering these expressions fairly accurate despite their simplicity. The quartic energy dependence highlights the flimsyness upon small twists ($\tau W \ll 1$), suggesting usage in sensitive torsion balances---torsion constant is practically zero. As soon as $\tau W \sim 1$, however, twisting rapidly becomes heavy.

Hydrogen-passivated ribbons have no stress at the edge, but for ribbons with a compressive edge stress the situation is more involved. The inset in Fig.~\ref{fig:energy}b shows how twisting makes the energy for unpassivated AGNRs even to decrease, inferring spontaneous twisting. This phenomenon, which has been reported before,\cite{bets_NR_09,koskinen_PRL_10} is susceptible to edge passivation and reconstruction.\cite{koskinen_PRL_08,koskinen_PRB_09,bhuang_PRL_09} In spontaneous twisting, however, energies are two orders of magnitude smaller than in forced twisting.

\begin{figure}[tb!]
\includegraphics[width=7cm]{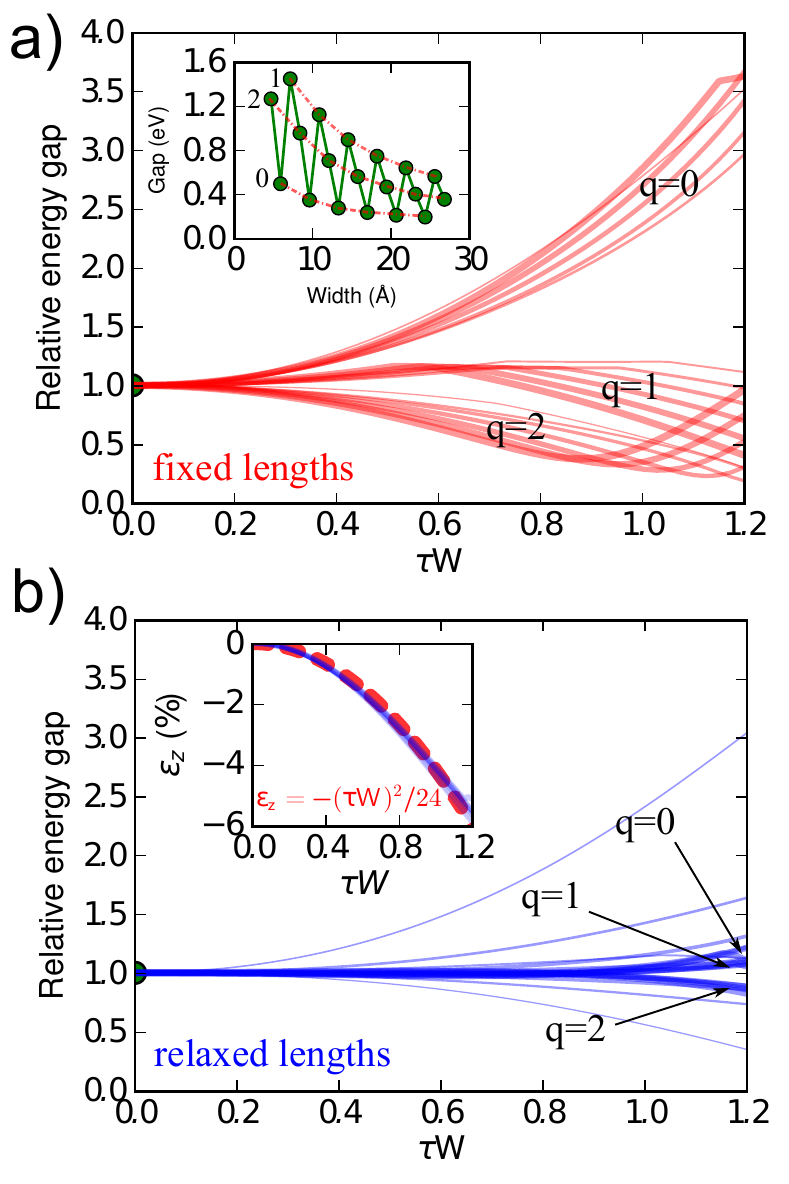}
\caption{(Color online) a) Relative changes in energy gaps of unpassivated, fixed-length $N$-AGNRs upon twisting. Inset: Absolute gaps for the corresponding flat ribbons; the $q$-families are defined in Eq.(\ref{eq:q}). b) As above, but with relaxed length. Inset: Axial strains that result from length relaxations follow the simple analytical expression. Line widths are proportional to ribbon widths.}
\label{fig:gaps}
\end{figure}

Let us now turn attention to the results on electronic structure. Figure~\ref{fig:gaps}a shows how the gaps of fixed-length, unpassivated $N$-AGNRs change upon twisting. Trends in the changes, just as in the gaps themselves (inset of Fig.~\ref{fig:gaps}a), fall into three families that I define here as
\begin{equation}
q = \text{mod}(N,3).
\label{eq:q}
\end{equation}
These trends agree with density-functional calculations of both finite and infinite ribbons.\cite{hod_NL_09, gunlycke_NL_10} What is fascinating, though, is that for relaxed lengths these trends nearly vanish (Fig.~\ref{fig:gaps}b). The three families are barely recognizable, substantial changes occurring only for the narrowest ribbons with $W\lesssim 1$~nm. The immediate question arises: how can one understand these differences?

\begin{figure*}[t!]
\includegraphics[width=16cm]{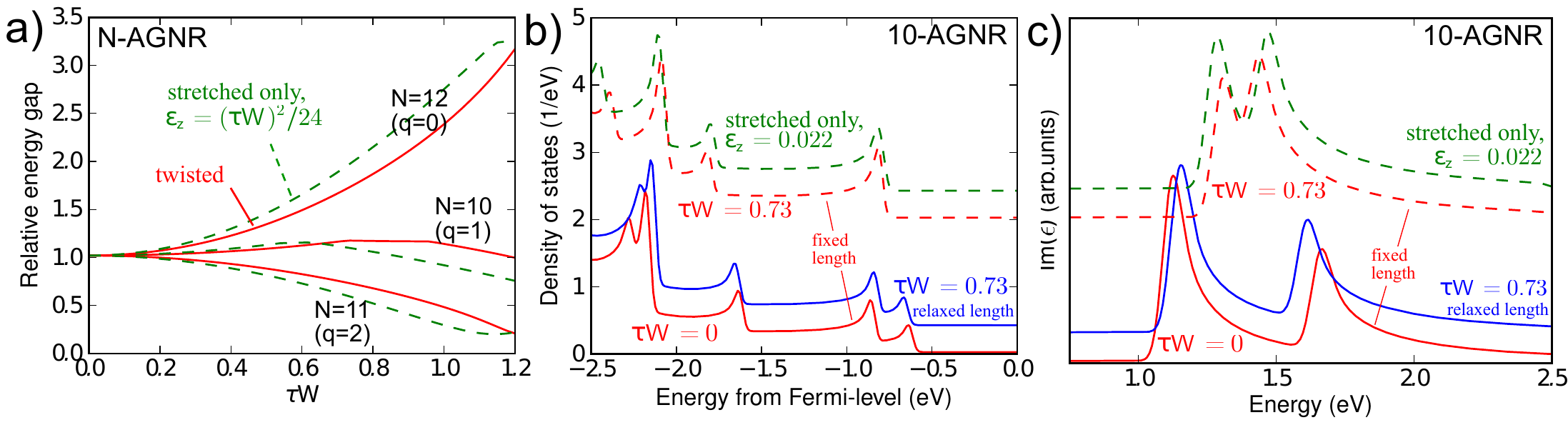}
\caption{(Color online) a) Relative changes in the energy gaps of three unpassivated $N$-AGNRs, that were being either twisted with fixed length or only stretched. b) Density of states for a $10$-AGNR that either was flat, was twisted to $\tau W=0.73$ with relaxed length, was twisted to $\tau W=0.73$ with fixed length, or was only stretched to $\varepsilon_z=(0.73)^2/24=0.022$ (with $\tau=0$). b) Imaginary part of the dielectric function under the same conditions as in b-panel. The dielectric function was calculated using random-phase approximation and polarization parallel to ribbon axis.}
\label{fig:origin}
\end{figure*}

As it will turn out, the key for understanding these trends is Eq.(\ref{eq:strain}). Fixed-length ribbons have $\varepsilon_z=0$, and yield average strains of $\varepsilon_{\text{avg}}=(\tau W)^2/24$ across the ribbons. Now, Fig.~\ref{fig:origin}a compares three AGNRs that either are twisted with fixed length or are strained with $\varepsilon_z=\varepsilon_{\text{avg}}=(\tau W)^2/24$ and $\tau=0$. The resulting similarity suggests the following interpretation: \emph{twisting by $\tau W$ can be imitated through stretching by $(\tau W)^2/24$}. The agreement is not perfect, and while it certainly could be improved by fitting, it would serve only little purpose, and merely give awkward dependencies on edge chiralities and passivations.

Simple analytical expressions are also more powerful than dummy fitting parameters. Namely, combined with an axial strain $\varepsilon_z$, the average strain from Eq.(\ref{eq:strain}) is
\begin{equation}
\varepsilon_{\text{avg}}=\varepsilon_z + (\tau W)^2/24.
\label{eq:epsavg}
\end{equation}
Minimizing the average strain $|\varepsilon_{\text{avg}}|$ with respect to $\varepsilon_z$ for given $\tau W$ minimizes also the total energy, inferring $\varepsilon_z=-(\tau W)^2/24$ and the trivial $\varepsilon_{\text{avg}}=0$. This provides a simple explanation to why energy gaps don't change in Fig.~\ref{fig:gaps}b: with relaxed lengths the \emph{average strains across the ribbons are zero}.

The inset of Fig.~\ref{fig:gaps}b shows that the relaxed length indeed follows the analytical estimate accurately. With this estimate the position-dependent strain becomes $\varepsilon(r)=\tau^2(r^2/2-W^2/24)$---the strain is compressive at the axis, zero at $r\approx 0.29 \cdot W$, and $(\tau W)^2/12$ at the edge. Length relaxation hence decreases the edge strain by two-thirds. The simulations agree surprisingly well with these simple expressions, especially in view of the noticeable anharmonic effects due to $\gtrsim 10$~\%\ strains at the edge.


The electronic structure is equivalent also beyond energy gaps. Fig.~\ref{fig:origin}b compares the density of states and Fig.~\ref{fig:origin}c the imaginary part of the dielectric function, both for a $10$-AGNR under four different conditions. In both figures a flat ribbon and a relaxed-length twisted ribbon, as well as a fixed-length twisted ribbon and a stretched flat ribbon share the same features among themselves; the trend-determining factor is the average strain.

Since most of the strain is at the edge, it is justified to ask whether twisting will bring about new localized electronic states at the edges. Closer analysis of the wave functions reveals, however, that no such localization takes place---also the the famous ZGNR edge states remained as before.\cite{nakada_PRB_96} The changes in electronic structure, induced by twisting, proved to be smooth.

To conclude, I have shown that changes in the electronic structure of graphene nanoribbons upon twisting are determined by average strain. Because of the simplicity of this cause---although not all the results were displayed---same reasoning applies irrespective of ribbon chirality or edge passivation. Similar stretching-induced effects were recently reported also in carbon nanotubes under pure bending.\cite{koskinen_PRB_10}  Since flat, stretched ribbons have been investigated more, and since Eq.~(\ref{eq:epsavg}) provides a direct mapping from stretched and twisted ribbons to stretched and flat ribbons, we thus have simple but powerful way to investigate ribbons under realistic experimental conditions. One should, however, be cautious not to take the twisting-stretching analogy too far---some properties, such as the ones related to magnetic flux,\cite{prada_PRL_10} may become fundamentally modified due to the uneven strain and the twisted geometry.

I acknowledge Oleg Kit for discussions and comments, the Academy of Finland for funding and the Finnish IT Center for Science (CSC) for computational resources.


\begin{thebibliography}{25}
\bibitem[{tau()}]{tau_note}
\bibinfo{note}{The parameter $\tau$ is defined with respect to the relaxed
  length of the flat ribbon.}
\end{thebibliography}

\end{document}